\documentclass{svjour3}
\usepackage{graphicx}
\usepackage{natbib}
\bibpunct{(}{)}{;}{a}{}{,}

\usepackage{epsfig}

\usepackage{amsmath,amssymb}      

\newcommand{\be}{\begin{equation}}
\newcommand{\ee}{\end{equation}}
\newcommand{\beq}{\begin{eqnarray}}
\newcommand{\eeq}{\end{eqnarray}}
\def\alf{Alfv\'en~}

\def \degmark{^\circ}

\newcommand{\ion}[2]{#1\,{\sc{#2}}}

\journalname{SSRv}

\begin{document}

\title{Equilibration processes in the Warm-Hot Intergalactic Medium}

\author{A.M. Bykov \and
        F.B.S. Paerels \and
        V. Petrosian}

\institute{A.M. Bykov \at A.F. Ioffe Institute of Physics and
Technology, St. Petersburg,
           194021, Russia \\
\email{byk@astro.ioffe.ru} \and
F.B.S. Paerels \at Columbia Astrophysics Laboratory, Department of
Astronomy, Columbia University, 550 West 120th Street, New York, NY
10027\\
\email{frits@astro.columbia.edu} \and
           V. Petrosian \at Department of Physics, Stanford University,
           Stanford, CA 94305\\
\email{vahe@astronomy.stanford.edu}
             }

\date{Received: 17 September 2007; Accepted: 30 October 2007 }

\maketitle

\begin{abstract}
The Warm-Hot Intergalactic Medium (WHIM) is thought to contribute
about $40-50$~\% to the baryonic budget at the present evolution
stage of the universe. The observed large scale structure is likely
to be due to gravitational growth of density fluctuations in the
post-inflation era. The evolving cosmic web is governed by
non-linear gravitational growth of the initially weak density
fluctuations in the dark energy dominated cosmology. Non-linear
structure formation, accretion and merging processes, star forming
and AGN activity produce gas shocks in the WHIM. Shock waves are
converting a fraction of the gravitation power to thermal and
non-thermal emission of baryonic/leptonic matter. They provide the
most likely way to power the luminous matter in the WHIM. The plasma
shocks in the WHIM are expected to be collisionless. 
Collisionless shocks produce a highly non-equilibrium state with
anisotropic temperatures and a large differences in ion and electron
temperatures. We discuss the ion and electron heating by the
collisionless shocks and then review the plasma processes
responsible for  the Coulomb equilibration and collisional
ionisation equilibrium of oxygen ions in the WHIM. MHD-turbulence
produced by the strong collisionless shocks could provide a sizeable
non-thermal contribution to the observed Doppler parameter of the UV
line spectra of the WHIM.
\end{abstract}

\section{Introduction}
\label{Introduction} Cosmological simulations of the large-scale
structure (LSS) predict that about $40-50$~\% of baryons at epoch $z
< 2$ could reside in the Warm-Hot Intergalactic Medium (WHIM) with
temperatures $10^5-10^7$ K at moderate overdensities $\delta <
100$ \citep{CenO99,Dave_ea01,Fang_ea02}. The
WHIM heating is due to shocks driven by gravitationally accelerated
flows in the LSS structure formation scenario (e.g.
\citealt{Kang_ea07}). Numerical simulations predict the observational
signatures of the web gas as a function of redshift. The simulations
account for feedback interactions between galaxies and the
intergalactic medium, and demonstrate that the X-ray and ultraviolet
\ion{O}{vi}, \ion{O}{vii} and \ion{O}{viii} lines and the \ion{H}{i} Lyman alpha line
are good tracers of low-density cosmic web filamentary structures
(e.g. \citealt{Hellsten_ea98,Tripp_ea00,Cen_ea01,Furlanetto_ea04}). Intervening metal absorption systems of
highly ionised C, N, O, Ne in the soft X-ray spectra of bright Active Galactic Nuclei (AGN) 
were suggested to be tracer of the WHIM. The predicted distribution
of ion column densities in the WHIM absorbers is steep enough to
provide  only a few systems with $N_{\mbox{O\,{\scriptsize VII}}}
> 10^{15}$~cm$^{-2}$  along an arbitrary chosen line of sight (e.g. \citealt{Fang_ea02}).
Therefore, the detection of the WHIM is particularly difficult and
requires very sensitive UV and X-ray detectors, both for absorption
and for emission processes (e.g. \citealt{Lehner_ea07,Nicastro_ea05,Kaastra_ea06,Takei_ea07} and
\citealt{richter2008,durret2008} - Chapters 3 and 4, this volume). Future projects and namely {\sl Cosmic Origin
Spectrograph} ({\sl COS}), the {\sl X-Ray Evolving Universe Spectrometer}
 ({\sl XEUS}), {\sl Constellation-X} and
  the {\sl Diffuse Intergalactic Oxygen Surveyor} ({\sl DIOS})
 will increase the signal-to-noise ratio in the spectra
allowing to study weak systems with $N_{\mbox{H\,{\scriptsize I}}} < 10^{12.5}$~cm$^{-2}$
and $N_{\mbox{O\,{\scriptsize VII}}} < 10^{15}$~cm$^{-2}$. Simulations of spectra of the
broad Ly$\alpha$ absorption lines and the highly ionised oxygen
lines in the weak systems require thorough modelling of physical
condition in the plasma (see e.g. \citealt{Mewe90,PaerelsK03,Kawahara_ea06}). We discuss in this paper the heating and
equilibration processes in the shocked WHIM plasma affecting the
spectral simulations. A discussion of collisionless shock physics
relevant to cosmological shocks in the WHIM can be found in \citealt{bykov2008} - Chapter 7, this volume. To this end, in this paper
we discuss first some specific features of collisionless shock
heating of ions of different charge states providing highly
non-equilibrium initial states just behind the magnetic ramp region
that relaxes to an equilibrium state through Coulomb collisions, and the
relation of the processes to simulations of emission/absorption
spectra of the WHIM and observational data analysis.

\section{WHIM heating and ion temperature evolution}

The plasma ion heating in the WHIM is most likely due to
cosmological shocks. The \alf Mach number of a shock propagating
through an ionised gas of local overdensity $\delta =
\rho/\left< \rho \right>$ at the epoch $z$ in the standard
$\Lambda$CDM cosmology is determined by
\begin{equation}
{\cal M}_{\rm a} = v_{\rm sh} (4\pi \rho_{\rm i})^{1/2}/B \approx
20.6\, v_{\rm s7} \delta^{1/2}  (1+z)^{3/2} 
(\Omega_{\rm b}h^2/0.02)^{1/2} B_{-9}^{-1},
\end{equation}
where $B_{-9}$ is the magnetic field just before the shock, measured
in nG and $v_{\rm s7}$ is the shock velocity in 10$^7$~cm\,s$^{-1}$, and
$\left< \rho \right>$ is the average density in the Universe.

The sonic Mach number for a shock propagating in a plasma of 
standard cosmic abundance is
\begin{equation}
{\cal M}_{\rm s} \approx 8.5\, v_{\rm s7}\,  [T_4  (1 +
f_{\rm ei})]^{-1/2},
\end{equation}
where $T_4$ is the plasma ion temperature measured in 10$^4$ K
(typical for a preshock photoionised plasma) and $f_{\rm ei} =
T_{\rm e}/T_{\rm i}$. An important plasma parameter is 
$$\beta = {\cal
M}^2_{\rm a}/{\cal M}^2_{\rm s} \approx 6 \delta 
(1+z)^{3}  (\Omega_{\rm b}h^2/0.02) B_{-9}^{-2} [T_4
 (1 + f_{\rm ei})].$$ 
It is the ratio of the thermal and
magnetic pressures. In hot X-ray clusters of galaxies the beta
parameter is $\sim$ 100 for  $\sim \mu$G magnetic fields in the
clusters. The most uncertain parameter is the magnetic field value
in the WHIM allowing for both $\beta \sim$ 1 and   $\beta \gg$ 1
cases.

In a supercritical collisionless shock the conversion of kinetic
energy of an initially cold flow to the ion distribution with  high
kinetic temperature occurs in the thin ion viscous jump. The width
of the ion viscous jump $\Delta_{\rm vi}$ in a collisionless shock
propagating through a plasma with $\beta \sim$ 1 is typically of the
order of a ten to a hundred times of the ion inertial length $l_{\rm
i}$ defined
 as $l_i = c/\omega_{\rm pi} \approx 2.3 \times 10^7 n^{-0.5}$ cm.
 Here $\omega_{\rm pi}$ is the ion plasma frequency. The ion inertial length in the WHIM can
 be estimated as 
$$l_{\rm i}
\approx 5.1 \times 10^{10}  \delta^{-1/2}  (1+z)^{-3/2} (\Omega_{\rm b}h^2/0.02)^{-1/2}\  {\mathrm cm}.$$
The width of the shock
transition region for magnetic field is also $\gtrsim 10 l_{\rm i}$
for a quasi-perpendicular shock, but it is often about ten times
wider for quasi-parallel shocks.

Properties of nonrelativistic shocks in a hot, low magnetised
plasma with high $\beta \gg$ 1 are yet poorly studied.
Measurements from the {\sl ISEE\,1} and {\sl ISEE\,2} spacecrafts
 were used by \citet{Farris_ea92} to examine the terrestrial bow shock
under high beta conditions. These measurements were compared with
and found to be in agreement with the predicted values of the
Rankine-Hugoniot relations using the simple adiabatic approximation
and a ratio of specific heats, $\gamma$, of 5/3. Large magnetic field
and density fluctuations were observed, but average downstream
plasma conditions well away from the shock were relatively steady,
near the predicted Rankine-Hugoniot values. The magnetic
disturbances persisted well downstream and a hot, dense ion beam was
detected leaking from the downstream region of the shock. The
observation proved the existence of collisionless shocks in high
beta plasma, but a detailed study of high beta shock structure is
needed for cosmological plasmas.

We discuss in the next section the ion heating in collisionless
shocks illustrating the most important features of the process
with the results of a hybrid simulation of the oxygen ions heating
in a quasi-perpendicular shock considered earlier by 
\citealt{bykov2008} - Chapter 7, this volume.

\subsection{Collisionless shock heating of the ions}

\begin{figure}[!ht]
\begin{tabular}{c}
\includegraphics[width=\textwidth]{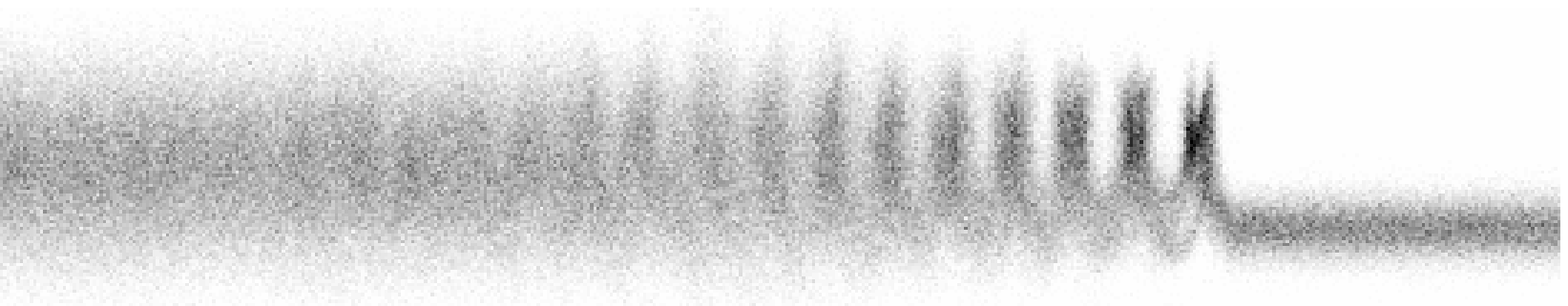}\\
\includegraphics[width=\textwidth]{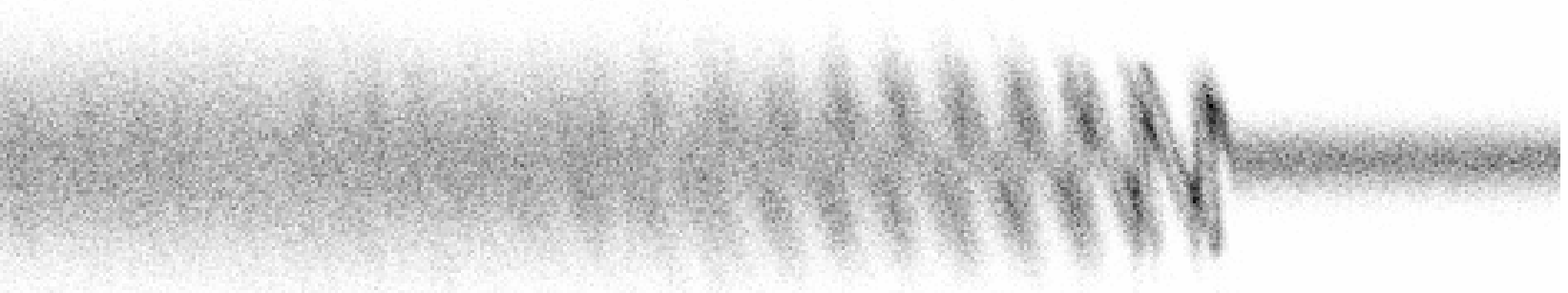} \\
\includegraphics[width=\textwidth]{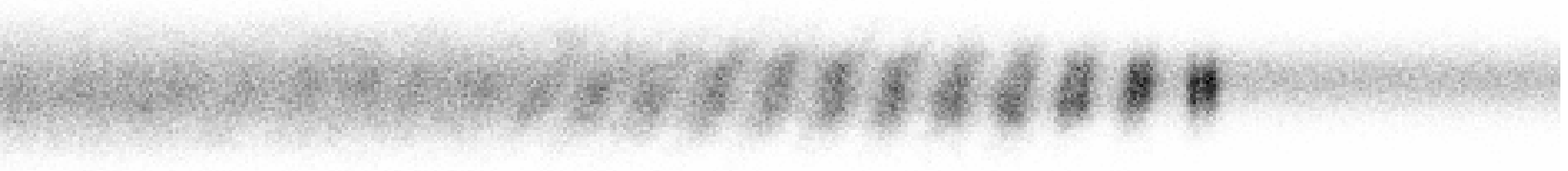} \\
\end{tabular}
\caption{\ion{O}{vii} phase density in a hybrid simulated
quasi-perpendicular shock (80$\degmark$ inclination). The shock is
moving from left to right in the reference frame where the
particle reflecting wall is at rest. The figures show the oxygen
phase densities in ${x - v_x}$, ${x - v_y}$ and ${x - v_z}$
projections from top to bottom respectively. The size of the
simulation box in x-dimension is about 300 $l_{\rm i}$.}
 \label{phase_ox}
\end{figure}

\begin{figure}    
\begin{center}
\hbox{
\psfig{file=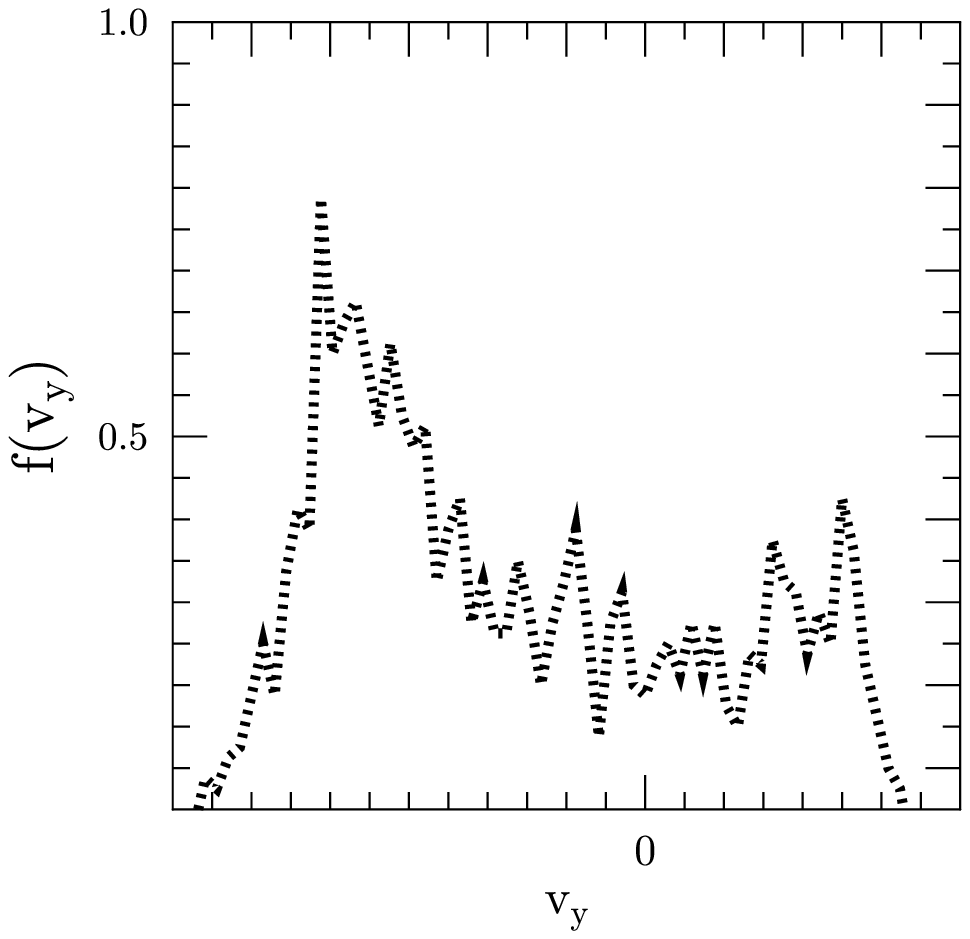,width=0.49\textwidth, bb=30 30 320 350,
clip} \psfig{file=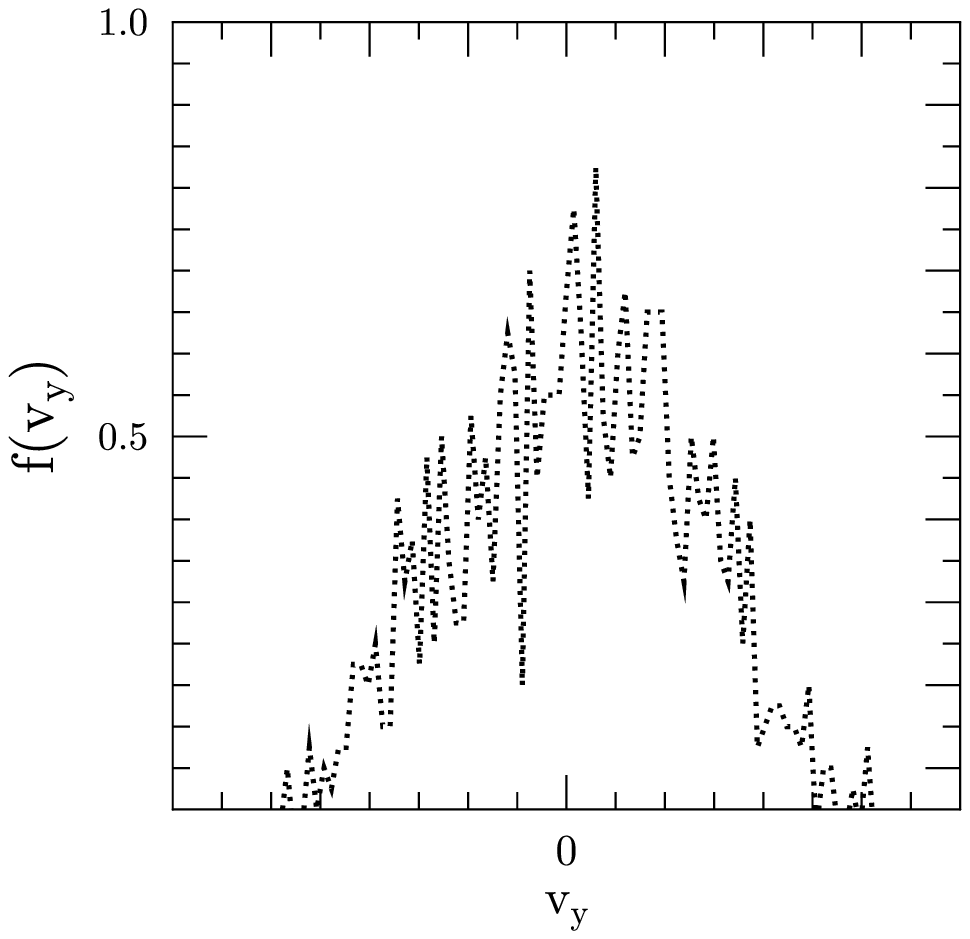,width=0.49\textwidth, bb=30 30 320
350 ,clip }}
\caption{Hybrid simulated  \ion{O}{vii} distribution function
(normalised) as a function of a random velocity component $\delta
v_{y} = v_{y} - <v_{y}>$ transverse to the downstream
magnetic field in a quasi-perpendicular shock (80$\degmark$
inclination). The shock propagates along the $x$-axis, while the
initial regular magnetic field is in the $x$--$z$ plane. In the left
panel the distribution in the viscous velocity jump is shown. The
right panel shows the distribution behind the jump at the position
of the left end in Fig.~\ref{phase_ox}. Multi-velocity structure of
the flow is clearly seen in the left panel, while it is relaxing to
quasi-Maxwellian in the right panel.} \label{PDF_O}
\end{center}
\end{figure}

Ion heating mechanisms in  collisionless shocks depend on the
shock \alf Mach number, the magnetic field inclination angle
($\theta_{\rm n}$), plasma parameter $\beta$ and the composition of
the incoming plasma flow. The structure of a supercritical shock is
governed by the ion flows instabilities (see e.g.
\citealt{Kennel_ea85,Lembege_ea04,Burgess_ea05}). In
a quasi-parallel shock ($\theta_{\rm n} \lesssim 45\degmark$) a mixed
effect of a sizeable backstreaming ion fraction and the ions
scattered by the strong magnetic field fluctuations (filling the
wide shock transition region) results in the heating of ions in the
downstream region. The ions reflected and slowed down by an electric
potential jump $\delta \phi$ at the shock ramp of a
quasi-perpendicular ($\theta_{\rm n} \gtrsim 45\degmark$) shock
constitute a multi-stream distribution just behind a relatively thin
magnetic ramp as it is seen in Fig.~\ref{phase_ox} and
Fig.~\ref{PDF_O} (left panel). The \ion{O}{vii} phase
densities and distribution functions were simulated with a hybrid
code for a quasi-perpendicular ($\theta_{\rm n} = 80\degmark$) shock
in a hydrogen-helium dominated plasma (see \citealt{bykov2008} - Chapter 7, this
volume). Phase densities {$x-v_x$,~ $x- v_y$,~ $x- v_z$} of
the \ion{O}{vii} ion are shown in Fig.~\ref{phase_ox} in the reference frame
where the particle reflecting wall (at far right) is at rest and the
shock is moving. The shock is propagating along the $x$ -axis from
the left to the right and the magnetic field is in the $x$--$z$
plane. The system is periodic in the $y$ dimension. The incoming
plasma beam in the simulation was composed of  protons (90~\%),
alpha particles (9.9~\%) and a dynamically insignificant fraction of
 oxygen ions (\ion{O}{vii}) with the upstream plasma parameter $\beta
\sim$ 1. The ions do not change their initial charge states in a few
gyro-periods while crossing the cosmological shock ramp where the
Coulomb interactions are negligible.

The simulated data in Fig.~\ref{phase_ox} show the ion velocities
phase mixing resulting in a thermal-like broad ion distribution at
a distance of some hundreds of ion inertial lengths in the shock
downstream (see the
 right panel in Fig.~\ref{PDF_O}). It is also clear in Fig.~\ref{phase_ox}
 that the shocked ion distribution tends to have anisotropy of the effective
 temperature. The temperature anisotropy
 $T_{\rm \perp} \sim 3\, T_{\rm \parallel}$ relative to the magnetic field was found in that simulation.
 Moreover, the hybrid simulation shows that the  $T_{\rm \perp}$ of
 the \ion{O}{vii} is about 25 times higher than the effective
 perpendicular temperature of the protons. Thus the ion downstream temperature declines from the linear
 dependence on the ion mass. The simulations show excessive heating of heavy ions in comparison with protons.

\citet{LeeW00} proposed a simplified analytical model to estimate
the ion perpendicular temperature dependence on $Z/A$, where $m_{\rm
i} = A m_{\rm p}$. Specifically, the model predicts the ratio of the
ion gyration velocity $v_{\rm ig2}$ in the downstream of a
perpendicular shock ($\theta_{\rm n} \sim 90\degmark$)  to the
velocity of the incident ion in the shock upstream, $v_{\rm 1}$, 
\be
 \frac{v_{\rm ig2}}{v_{\rm 1}} = \left|\left(1 - \alpha \frac{Z}{A}\right)^{1/2} - \frac{B_{\rm t1}}{B_{\rm
 t2}}\right|, \label{eq:LW00}
\ee 
where $\alpha = 2\, {\mathrm e}\, \delta \phi /m_{\rm p}v_{\rm 1}^2 < 1$,
and the potential jump  $\delta \phi$ is calculated in the shock
normal frame (see \citealt{LeeW00}). The model is valid for the ions
with  gyroradii larger than the shock transition width $\Delta_{\rm
vi}$. It is not a fair approximation for the protons, but it is much better for
temperature estimation of heavy ions just behind the shock magnetic
ramp. The model of ion heating in the fast, supercritical
quasi-perpendicular ($\theta_{\rm n} \gtrsim 45\degmark$) shocks of
${\cal M}_{\rm a} \gtrsim 3$ predicts a higher downstream perpendicular
temperature for the ions with larger $A/Z$.


\subsection{Collisionless heating of the electrons}

The initial electron temperature just behind the viscous ion jump of
a cosmological shock depends on the collisionless heating of the
electrons. The only direct measurements of the electron heating by
collisionless shocks are those in the Heliosphere. The
interplanetary shock data compiled by \citet{Schwartz_ea88} show a
modest, though systematic departure of the electron heating from
that which would result from the approximately constant ratio of the
perpendicular temperature to the magnetic field strength (i.e.
adiabatic heating). Thus, some modest non-adiabatic electron
collisionless heating is likely present. In the case of a
nonradiative supernova shock propagating through {\it partially
ionised} interstellar medium the ratio $T_{\mathrm e}/T_{\mathrm i}$ in a thin layer
(typically $< 10^{17}$ cm) just behind a shock can be tested using
the optical diagnostics of broad and narrow Balmer lines (e.g.
\citealt{Raymond01}). High resolution {\sl Hubble Space Telescope  (HST)} Supernova remant (SNR) images make that
approach rather attractive. A simple scaling $T_{\rm e}/T_{\rm i}
\propto v_{\rm sh}^{-2}$ was suggested by \citet{Ghavamian_ea07} to
be consistent with the optical observations of SNRs.

Strong shocks are thought to transfer a sizeable fraction of the bulk
kinetic energy of the flow into large amplitude nonlinear waves in
the magnetic ramp region. The thermal electron velocities in the
ambient medium are higher than the shock speed if the shock Mach
number ${\cal M}_{\rm s} < \sqrt{m_{\mathrm p}/m_{\mathrm e}}$, allowing for
a nearly-isotropic angular distribution of the electrons. Non-resonant
interactions of these electrons with large-amplitude turbulent
fluctuations in the shock transition region could result in
collisionless heating and pre-acceleration of the electrons
\citep{BykovU99,Bykov05}. They calculated the electron
energy spectrum in the vicinity of the shock waves and showed that
the heating and pre-acceleration of the electrons occur on a scale
of the order of several hundred ion inertial lengths in the vicinity
of the viscous discontinuity. Although the electron distribution
function is in a significantly non-equilibrium state near the shock
front, its low energy part can be approximated by a Maxwellian
distribution. The effective electron temperature just behind the
front, obtained in this manner, increases with the shock wave
velocity as $T_{\rm e} \propto v_{\rm sh}^b$ with $b \leq 2$. They
also showed that if the electron transport in the shock transition
region is due to turbulent advection by strong vortex fluctuations
of the scale of about the ion inertial length, then the nonresonant
electron heating is rather slow (i.e. $b \leq 0.5$). The highly
developed vortex-type turbulence is expected to be present in the transition
regions of very strong shocks. That would imply that the initial
$T_{\rm e}/T_{\rm i} \propto v_{\rm sh}^{(b-a)}$ just behind the
transition region would decrease with the shock velocity for ${\cal
M}_{\rm s} \gg 1$. Here the index $a$ is defined by the relation
$T_{\rm i} \propto v_{\rm sh}^a$ for a strong shock. The degree of
electron-ion equilibration in a collisionless shock is a declining
function of shock speed. In the case of strong vortex-type
turbulence in the shock transition region one expects in the
standard ion heating case with $a=2$ and rather small $b$ to have
$(a - b) \lesssim $ 2. That $T_{\rm e}/T_{\rm i}$ scaling is somewhat
flatter, but roughly consistent with, that advocated by
\citet{Ghavamian_ea07}. On the other hand in a collisionless shock
of a moderate strength ${\cal M}_{\rm s} < 10$ the electron
transport through the magnetic ramp region could be diffusive,
rather than by the turbulent advection by a strong vortexes. That
results in a larger degree of the collisionless electron
heating/equilibration in the shocks as it is shown in Fig.~4 of the
paper by \citet{BykovU99}. Recently, \citet{MarkevitchV07} argued
for the collisionless heating/equilibration of the electron
temperature in the bow shock of ${\cal M}_{\rm s} \sim 3$ in the 1E~$0657-56$ cluster.

If the local Mach number ${\cal M}_{\rm s}$ of the incoming flow in
a strong shock wave exceeds $\sqrt{m_{\rm p}/m_{\rm e}}$, which
could occur in the cluster accretion shocks, the thermal electron
distribution becomes highly anisotropic and high frequency whistler
type mode generation effects could become important.
\citet{Levinson96} performed a detailed study of resonant electron
acceleration by the whistler mode for fast MHD shock waves. Electron
heating and Coulomb relaxation in the strong  accretion shocks in
clusters of galaxies was discussed in details by \citet{FoxL97}.

We summarise this section concluding that a collisionless shock
produces in the downstream flow a highly non-equilibrium plasma
state with strongly different temperatures of the electrons and
ions of different species. Moreover, the postshock ion
temperatures are anisotropic. The width of the collisionless shock
transition region is smaller by many orders of magnitude than the
Coulomb mean free path (that is of a kiloparsec range). We
consider now the structure and the processes in the postshock
Coulomb equilibration layers in the WHIM.

\section{Coulomb relaxation of temperatures in the WHIM}

\subsection{Relaxation of the ion temperature anisotropy}

A plasma flow partly randomised just behind the viscous ion jump
in a collisionless shock transition has an anisotropic velocity
distribution with respect to the mean magnetic field. In
Fig.~\ref{phase_ox}  the oxygen phase densities in the ${v_x - x}$
and ${v_y - x}$ projections transverse to the mean field have
a substantially wider distribution than that in the projection ${v_z
- x}$ parallel to the magnetic field. In many cases the ion
velocity distributions (like that in Fig.~\ref{PDF_O} right) can
be approximated with a quasi-Maxwellian distribution introducing
some effective kinetic temperature (more exactly it is the second
moment of the distribution). Then the parallel and perpendicular
(to the mean field) temperatures are different and we approximate
the 3D particle distribution as
\be f(v_{\rm \perp},v_{\rm \parallel}) = \left(\frac{m}{2\pi T_{\rm
\perp}}\right)\left(\frac{m}{2\pi T_{\rm
\parallel}}\right)^{1/2}\  \exp{\left(- \frac{m
v_{\rm \perp}^2}{2T_{\rm \perp}} -\frac{m v_{\rm
\parallel}^2}{2T_{\rm \parallel}}\right)}. \label{eq:anisMaxw} \ee

We will measure the temperatures in energy units in most of the
equations below (thus ${\mathrm k}_{\mathrm B} T \to T$). \citet{IchimaruR70} (see also
a comment by \citealt{Kaiser79}) obtained the following equation to
describe the ion anisotropy relaxation
\be \frac{{\mathrm d}T_{\rm \perp}}{{\mathrm d}t} = -\frac{1}{2} \frac{dT_{\rm
\parallel}}{dt} = -\frac{T_{\rm \perp} -T_{\rm \parallel}}{\tau_{\rm i}}.
\label{eq:anis1}\ee
Here 
\be \tau^{-1}_{\rm i} = \frac{8 \pi^{1/2} n_{\rm i} Z^2 {\mathrm e}^4 \ln
\Lambda}{15 m^{1/2}_{\rm i} T^{3/2}_{\rm eff}}, \label{eq:anis2} \ee
where $\ln \Lambda$ is the Coulomb logarithm and the effective ion
temperature is defined as 
\be \frac{1}{T^{3/2}_{\rm eff}} =
\frac{15}{4} \int^{1}_{-1} {\mathrm d}\mu \frac{\mu^2
(1-\mu^2)}{[(1-\mu^2)T_{\rm \perp} + \mu^2 T_{\rm
\parallel}]^{3/2}}. \label{eq:anis3}
\ee
These equations were obtained under the assumption that the electrons
have no dynamical role, but provide a static dielectric background
to the ions. We can directly apply Eqs.~\ref{eq:anis1}$-$\ref{eq:anis3} to isotropisation of the plasma field particles
(i.e. protons in our case). Isotropisation of the minor ion
components is mainly due to their interactions with protons and
helium because of a low metal number density in cosmic plasma.

\subsection{Relaxation of the ion and electron temperatures}
Following shock heating and the temperature isotropisation a
quasi-Maxwellian distribution will be established within all plasma
components, i.e. groups of identical particles, after a time scale
given by Eq.~\ref{eq:anis2}. The effective temperatures differs strongly
between the components. All the plasma particles will undergo
Coulomb collisions with the protons, alpha particles and electrons
dominating the WHIM plasma, resulting in the temperature
equilibration.  \citet{Spitzer62} found that the  temperature
relaxation of a test particle of type $a$ with plasma field
particles can be approximately estimated from
\be \frac{{\mathrm d}T_{\rm a}}{{\mathrm d}t} =  \frac{T_{\rm p} -T_{\rm a}}{\tau_{\rm
ap}} \ee
where 
\be \tau^{-1}_{\rm ap} = \frac{8 (2\pi)^{1/2}}{3}\ 
\frac{n_{\rm p} Z^2 e^4 \ln \Lambda}{m_{\rm a} m_{\rm p}}\ 
\left( \frac{T_{\rm p}}{m_{\rm p}} + \frac{T_{\rm a}}{m_{\rm
a}}\right)^{-3/2} \ee

In the thermal equilibrium state the postshock plasma must have a
single equilibrium temperature $T_{\rm eq}$.  In a fully ionised
plasma without energy exchange with external components (i.e.
radiation or plasma wave dissipation) $T_{\rm eq}$ can be found from
the condition of constant pressure in the plane shock downstream
resulting in
\be T_{\rm eq}  = \sum  n_{\rm a}T_{\rm a0}/\sum  n_{\rm a} \ee

\begin{figure}    
\begin{center}
\includegraphics[width=0.7\textwidth]{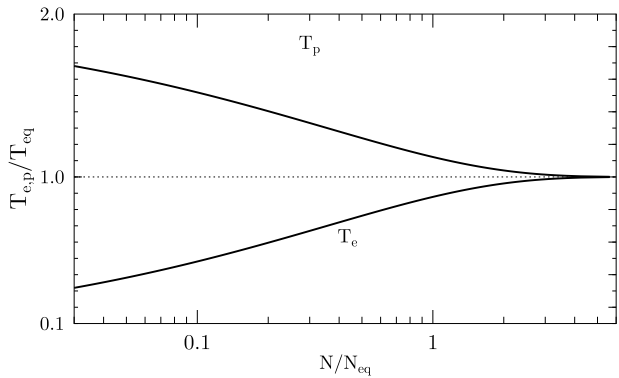}
\caption{Postshock temperature equilibration between the ion and
electron components due to the Coulomb interactions.}
\label{equil_ie}
\end{center}
\end{figure}

In cosmic plasmas it is often a fair approximation to estimate
$T_{\rm eq}$ from the equation $2\,T_{\rm eq} = T_{\rm e} + T_{\rm
p}$. Then following \citet{Sivukhin66} the charged particle
equilibration can be approximately described through the equation
\be \ln \left|\frac{\sqrt{T_{\rm e}} - \sqrt{T_{\rm
eq}}}{\sqrt{T_{\rm e}} + \sqrt{T_{\rm eq}}}\right| = -
\frac{t}{\tau_{\rm eq}} - \frac{2}{3}\left(\frac{T_{\rm e}}{T_{\rm
eq}}\right)^{3/2} - 2\left(\frac{T_{\rm e}}{T_{\rm eq}}\right)^{1/2}
+ C_{\rm e}, \label{eq:Trelax} \ee 
where $C_{\rm e}$ is a constant
to be determined from the initial temperature $T_{\rm a0}$ of a
relaxing component $a = {\mathrm e, p}$, 
\be \tau_{\rm eq}^{-1} \approx
\frac{16 (2\pi)^{1/2}}{3}\  \frac{n_{\rm p} e^4 \ln
\Lambda}{m_{\rm e} m_{\rm p}}\  \left( \frac{T_{\rm eq}}{m_{\rm
e}}\right)^{-3/2}. \label{eq:eq_tau} \ee
Eq.~\ref{eq:Trelax} and  Eq.~\ref{eq:eq_tau} allow to
calculate the structure of relaxation layers to be seen behind a
collisionless shock in the WHIM. In Fig.~\ref{equil_ie} we
illustrate the postshock equilibration of initially cold electrons
with the protons initially heated at the ion viscous jump of a
collisionless shock transition. The width of the ion viscous jump
in cosmological shocks is negligible compared with the
equilibration length $x_{\rm eq} = u_2 \tau_{\rm eq}$, where
$u_2$ is the downstream flow velocity in the shock rest frame. The
characteristic column density $N_{\rm eq} = n_2  x_{\rm eq}$
to be traversed by protons and electrons in the downstream plasma
(of a number density $n_2$) before the temperature equilibration,
as it follows from Eq.~\ref{eq:eq_tau}, does not depend on the
plasma number density, and $N_{\rm eq} \propto v_{\rm sh}^4$. The
corresponding shocked WHIM column density can be expressed through
the shock velocity $v_{\rm 7}$ measured in 100~km\,s$^{-1}$, assuming a
strong shock where $u_2 = v_{\rm sh}/4$:
$$
N_{\rm eq} \approx 5\times 10^{17}\  v_{\rm 7}^4 \ [\ln \Lambda]^{-1} \ {\mathrm{cm}}^{-2}, \\$$
or through the shocked WHIM gas temperature $T_{6}$ (measured in
10$^6$ K):
$$
N_{\rm eq} \approx 2.5\times 10^{17}\  T_{6}^2  [\ln \Lambda]^{-1} \ {\mathrm{cm}}^{-2}. \\$$

The Coulomb logarithm for the WHIM condition is $\ln \Lambda \sim$
40. It follows from Fig.~\ref{equil_ie} that in the Coulomb
relaxation model the postshock plasma column density $N_{\rm H} >
3 N_{\rm eq}$ ensures the equilibration at a level better than
1~\%.

The metal ions can be initially heated at the shock magnetic ramp to
high enough temperatures $> A T_{\rm p}$ (see e.g.
\citealt{Korreck_ea07}) for a recent analysis of interplanetary
collisionless shock observations with {\sl Advanced Composition
Explorer}). However, in a typical case the depth $N > 3 N_{\rm eq}$
is enough for the ion temperature equilibration.

To study UV and X-ray spectra of the weak systems  ($N_{\mbox{H\,{\scriptsize I}}} <
10^{12.5}$~cm$^{-2}$ and $N_{\mbox{O\,{\scriptsize VII}}} < 10^{15}$~cm$^{-2}$) modelling of
shocked filaments of $N_{\rm H} \lesssim 10^{17}$~cm$^{-2}$ would require an
account of non-equilibrium effects of low electron temperature
$T_{\rm e}/T_{\rm eq} < 1$.

\subsection{Effect of postshock plasma micro-turbulence on the line widths}

We consider above only the WHIM temperature evolution due to the
Coulomb equilibration processes. Shocks producing the WHIM could
propagate through inhomogeneous (e.g. clumpy) matter. The
interaction of a shock with the density inhomogeneities results in the
generation of MHD-waves (\alf and magnetosonic) in the shock
downstream (see e.g. \citealt{VBT93}). The MHD-wave dissipation in the
shock downstream could selectively heat ions, being a cause of
a non-equilibrium $T_{\rm e}/T_{\rm i}$ ratio. In case of a strong
collisionless shock propagating in a turbulent medium, cosmic ray
acceleration could generate a spectrum of strong MHD-fluctuations
(see e.g. \citealt{BlandfordE87,Bell04,Vladimirov_ea06}). These MHD-fluctuations could carry a
substantial fraction of the shock ram pressure to the upstream andthen to the downstream providing a heating source throughout the
downstream. The velocity fluctuations could also produce non-thermal
broadening of the lines. The amplitude of bulk velocity fluctuations
is about the \alf velocity since $\delta B \sim B$ in the shock
precursor. The Doppler parameter $b$ derived from high resolution UV
spectra of the WHIM (see e.g. \citealt{Lehner_ea07,richter2008} - Chapter 3, this volume):
$$
b^2 = \frac{2T}{m_a} + b_{\rm nt}^2
$$
 would have in the micro-turbulent limit a non-thermal contribution
 $$
 b_{\rm nt}^2 = \frac{2v_{\rm turb}^2}{3} = \frac{C_{\nu}
 B^2}
 {6\, \pi  \left< \rho \right>\, \delta}.
$$
The factor $C_{\nu}$ here accounts for the amplitude and spectral
shape of the turbulence. For a strong MHD-turbulence that was found
in the recent models of strong collisionless shocks with efficient
particle acceleration (e.g. \citealt{Vladimirov_ea06}) one can get
$C_{\nu}\sim 1$ in the WHIM, and then
 $$
 b_{\rm nt} \approx 4 B_{\rm
 -9} (\Omega_{\rm b}h^2/0.02)^{-1/2} (1+z)^{-3/2}
 \delta^{-1/2} \ \ {\mathrm{km}}\,{\mathrm s}^{-1}.
$$
The estimation of $b_{\rm nt}$ given above for a strong \alf
turbulence may be regarded as an upper limit for a system with a
modest MHD turbulence. The non-thermal Doppler parameter  does not
depend on the ion mass, but $b_{\rm nt} \propto \delta^{-1/2}
B(z)$. Thus the account of $b_{\rm nt}$ could be important for high
resolution spectroscopy of metal lines, especially if the strong
shocks can indeed highly amplify local magnetic fields. That is
still to be confirmed, but a recent high resolution observation of a
strong Balmer-dominated shock on the eastern side of Tycho's
supernova remnant with the {\sl Subaru Telescope} supports the
existence of a cosmic ray shock precursor where gas is heated and
accelerated ahead of the shock \citep{Lee_ea07}. High resolution UV
spectroscopy of the WHIM could allow to constrain the intergalactic
magnetic field. Internal shocks in hot X-ray clusters of galaxies
have modest Mach numbers and the effect of the \alf turbulence is
likely less prominent than that in the strong accretion shocks in
clusters and in the cosmic web filaments. X-ray line broadening by
large scale bulk motions in the hot intracluster medium was
discussed in detail by \citet{FoxL97} and \citet{InogamovS03}.

\begin{figure}
\begin{center}
\hbox{ \psfig{file=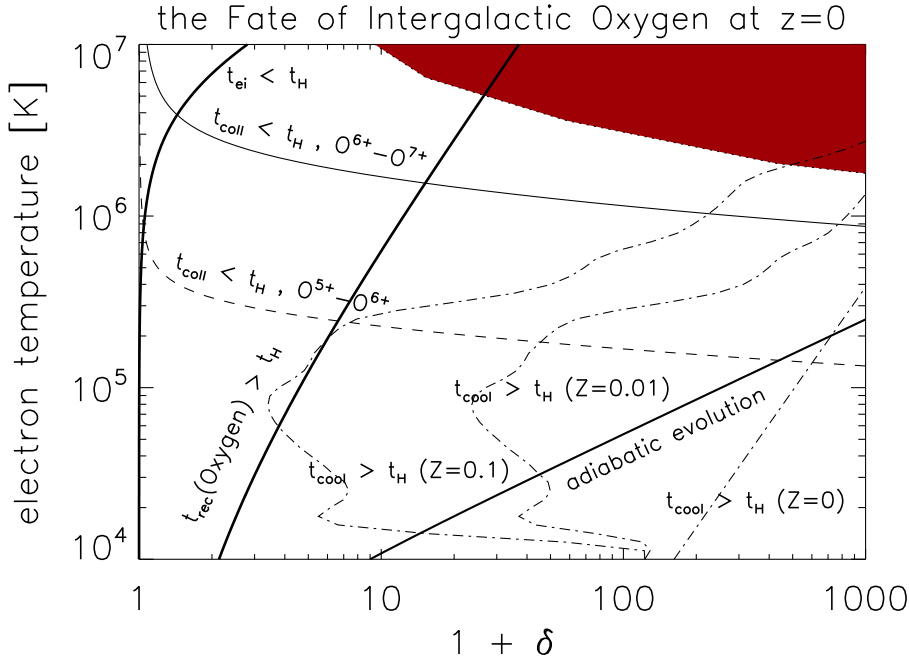,width=12.5cm,clip=}} \caption{Phase
diagram for oxygen in the IGM at redshift $z=0$. Density is
parameterised by the $\delta$ value as was defined above. The solid
line in the upper left hand corner labelled '$t_{\rm ei} < t_H$'
indicates where electron and proton fluids reach equilibrium
 in a Hubble time (t$_{\rm H}$). Low density gas will not radiatively cool over a Hubble
time to the left of the boundaries marked '$t_{\rm cool} > t_{\rm
H}$'. The three curves are labelled with the metallicity, $Z$,
expressed as a fraction of Solar metallicity.  The solid curve
labelled '{\it adiabatic evolution'} indicates the locus of gas that
has only undergone adiabatic compression or expansion since high
redshift (initial condition $T_{\mathrm e} \sim 10^4$ K).  Critical boundaries
for the ionisation equilibrium of oxygen are: the shaded area in the
upper right hand corner indicates the regime where the collisional
ionisation timescale is shorter than the photoionisation timescale,
for ionisation \ion{O}{viii} $\rightarrow$ \ion{O}{ix}. The two boundaries
labelled '$t_{\rm coll}  < t_{\rm H}$' indicate where the collisional
ionisation timescale becomes shorter than the Hubble time. At lower
temperature, the ionisation balance cannot be in (collisional)
equilibrium. Upper (solid) curve is for ionisation \ion{O}{vii} (O$^{+6}$)
$\rightarrow$ \ion{O}{viii} (O$^{+7}$), lower (dashed) curve for \ion{O}{vi} (O$^{+5}$)
$\rightarrow$ \ion{O}{vii} (O$^{+6}$). The steep solid curve labelled '$t_{\rm
rec}{\rm (Oxygen)} > t_{\rm H}$' indicates where the radiative
recombination timescale (\ion{O}{ix} $\rightarrow$ \ion{O}{viii}) exceeds
the Hubble time (no recombination at low densities). }
\label{charge_states}
\end{center}
\end{figure}

\section{Ionisation state of the WHIM}
To simulate absorption spectra of bright quasars in the intervening
WHIM filaments (e.g. \citet{Kawahara_ea06} as an example of such a
modelling) one should solve the ionisation balance equations for the
charge states of metal ions with account taken of all the LTE
processes and also the nonthermal particle contribution (see
\citet{Porquet_ea01} for a discussion of a role of non-relativistic
super-thermal distributions). The ionisation balance equation can be
written as
\begin{eqnarray}
\nonumber
  \dot{n}_{\rm q}&=&
   n_{\mathrm e}\left[n_{\rm q-1}C_{\rm q-1}-n_{\rm q}C_{\rm q}-n_{\rm q}\alpha_{\rm q}+n_{\rm q+1}\alpha_{\rm q+1}\right]+\\
\label{nidot}
               &&+\sum_{j={\mathrm{H,\,He,\,He^{+}}}}n_{\rm j}\left[n_{\rm q-1}V^{\rm ion}_{\rm j,q-1}-\nonumber
                   n_{\rm q}(V^{\rm rec}_{\rm j,q}+V^{\rm ion}_{\rm j,q})+n_{\rm q+1}V^{\rm rec}_{\rm j,q+1}\right]+\\
\nonumber
               &&+n_{\rm q-1}R_{\rm q-1}-n_{\rm q}R_{\rm q}.
\end{eqnarray}
Here $q$  is the charge state of an element, $C_{\rm q}$  is the
 collisional (and autoionisation) rate $q\to q+1$,
$\alpha_{\rm q}$  are the radiative and the dielectronic ionisation
rates $q\to q-1$ (in cm$^{3}$\,s$^{-1}$), $V^{\rm rec}_{\rm j,q}$.
$V^{\rm ion}_{\rm j,q}$ are the charge exchange rates with the ion
$j$ (in cm$^{3}$\,s$^{-1}$) and  $R_{\rm q}$ is the photoionisation
rate of an ion (in s$^{-1}$). The rates of different processes can
be calculated for different temperature regimes (see \citealt{kaastra2008} - Chapter 9, this volume). We just limit our discussion here to
one example of such a simulation.

In Fig.~\ref{charge_states} we illustrate {\sl collisional
ionisation equilibrium} curves of oxygen ions in the present epoch
(at $z=0$) as a function of WHIM  density $\delta$. The various
boundaries separate regimes under which a certain process does, or
does not attain equilibrium over a Hubble time. We show in
Fig.~\ref{charge_states} some critical boundaries for kinetic and
thermal equilibrium. The solid line in the upper left hand corner
labelled 't$_{\rm ei} <$ t$_{\rm H}$' indicates where electron and
proton fluids reach kinetic equilibrium (proton temperature equal to
electron temperature) in a Hubble time (t$_{\rm H}$): at low density
and high temperature, such equilibrium does not attain. Low density
gas will not radiatively cool over a Hubble time to the left of the
boundaries marked '$t_{\rm cool} >  t_{\rm H}$'. The cooling time
was calculated for collisionally ionised gas. The solid curve
labelled '{\it adiabatic evolution'} indicates the locus of gas that
has only undergone adiabatic compression or expansion since high
redshift (initial condition $T_{\mathrm e} \sim 10^4$ K); all shock-heated gas
will be above this line right after passing through a shock. The
shaded area in the upper right hand corner indicates the regime
where the collisional ionisation timescale is shorter than the
photoionisation timescale, for ionisation \ion{O}{viii} $\rightarrow$
\ion{O}{ix}. The two boundaries labelled '$t_{\rm coll}  < t_{\rm H}$'
indicate where the collisional ionisation timescale becomes shorter
than the Hubble time. At lower temperature, the ionisation balance
cannot be in (collisional) equilibrium. . The steep solid curve labelled
'$t_{\rm rec}{\rm (Oxygen)} > t_{\rm H}$' indicates where the
radiative recombination timescale (\ion{O}{ix} $\rightarrow$ \ion{O}{viii})
exceeds the Hubble time since there is no recombination at low
densities.

Note that we illustrate here only the collisional
equilibrium case. For the more appropriate case of radiative cooling
in photoionisation equilibrium, the cooling times will be even
longer, due to the fact that a photoionised plasma is highly
overionised compared to the characteristic ionisation- and
excitation potentials, which suppresses the (very effective)  
collisional cooling contribution.
In our figure, the lines
t$_{\rm cool}$ = t$_{\rm H}$ will shift to the right if we calculate
with the probably more realistic case of photoionisation equilibrium.
However, as we argued above the collisional equilibrium is also of
interest, since it represents a conservative case. The range of
ionisation states of oxygen in the WHIM filaments can be observed in
the absorption spectra of bright quasars.

\section{Conclusions}

We discussed some specific features of the WHIM heating processes by
collisionless plasma shocks driven by gravitationally accelerated
flows in the LSS structure formation scenario.

$\bullet$ Collisionless shock heating of ions in the WHIM results in
a highly non-equilibrium initial state with a strongly anisotropic
quasi-Maxwellian ion distribution just behind a very thin magnetic
ramp region. The ion temperatures  just behind the shock depend on
the ion atomic weight, the charge state and the shock magnetic field
inclination. The ion temperatures could decline from a simple linear
scaling with the ion mass providing an excessive heating of heavy
ions.

$\bullet$ The ion and electron temperatures relax to the equilibrium
state through Coulomb collisions in a layer of the depth $N_{\rm eq}
\sim 10^{16} v_{\rm 7}^4$~cm$^{-2}$ behind a shock of velocity
$v_{\rm 7}$.

$\bullet$ Strong collisionless plasma shocks with an efficient Fermi
acceleration of energetic particles can generate strong MHD waves
in the downstream region that will result in a non-thermal
broadening of the emission/absorption lines in the observed WHIM
spectra.

\begin{acknowledgements} 
The authors thank ISSI (Bern) for support of
the team ``Non-virialized X-ray components in clusters of galaxies''.
    A.M.B. thanks M. Yu. Gustov for his help with hybrid shock simulations,
    he acknowledges the RBRF grant 06-02-16844
    and a support from RAS Presidium Programs. A support from
    NASA ATP (NNX07AG79G) is  acknowledged.
\end{acknowledgements}

\bibliographystyle{aa}
\bibliography{08_bykov}
\end{document}